\documentclass[conference]{IEEEtran}
\IEEEoverridecommandlockouts
\usepackage{placeins}
\usepackage{cite}
\usepackage{amsmath,amssymb,amsfonts}
\usepackage{graphicx}
\usepackage{textcomp}
\usepackage{xcolor}  
\usepackage{booktabs} 
\usepackage{multirow}   
\usepackage{bm}
\usepackage[ruled,vlined,linesnumbered]{algorithm2e}
\usepackage[table]{xcolor}
\def\BibTeX{{\rm B\kern-.05em{\sc i\kern-.025em b}\kern-.08em
    T\kern-.1667em\lower.7ex\hbox{E}\kern-.125emX}}

\usepackage{comment}

\begin{document}
\title{FedCFM: Federated Continual Domain Generalization for Fake Speech Detection via Conditional Flow Matching\\
	\thanks{This work was supported in part by 2025 Telecom	NJUPT Industry-University-Research Cooperation Project under Contract JSSGS2500832EGN00, and in part by the National Natural Science Foundation of China under Grants 62071242 and 62601428.\\
	\textsuperscript{*} Corresponding author}
	}

\author{
	\IEEEauthorblockN{Yingjian Yu, Haiyan Guo\textsuperscript{*}, Tianshun Wang, Xinzhou Xu, Zirui Ge, Chi Liu, Ziheng Liu} 
	\IEEEauthorblockA{School of Communications and Information Engineering,\\
		Nanjing University of Posts and Telecommunications, P.R. China} 
}

\maketitle

\begin{abstract}
The generalization ability of Fake Speech Detection (FSD) models is crucial for real-world deployment. Existing multi-dataset co-training methods rely on fixed training sets and cannot adapt to emerging spoofing types. Although continual learning has been explored, many approaches overlook limited data storage at individual devices, thereby restricting practical applicability. To address this, we propose FedCFM, a Federated continual domain generalization framework via Conditional Flow Matching (CFM) for collaboration without sharing raw speech data across distributed clients facing diverse and evolving spoofing attacks. Each client trains a CFM-based generator to model spoof-type-specific embedding distributions, and cross-client generator exchange enables synthesis of unseen spoof-type embeddings for continual classifier updating through generative replay and knowledge distillation. With the same training datasets, FedCFM achieves lower EER than the evaluated centralized and federated domain generalization baselines, demonstrating strong cross-domain generalization. Code will be released on https://github.com/jspycpp/FedCFM.
\end{abstract}


\section{Introduction}
Advances in speech synthesis have diversified spoofing attacks and increased distribution shifts across attack types. Representative FSD methods, including AASIST~\cite{b1}, RawNet2~\cite{b2}, RawBMamba~\cite{b3}, and self-supervised approaches~\cite{b4,b5,b6}, achieve competitive performance but are often trained on a single dataset, limiting generalization to unseen attacks, speakers, languages, and recording conditions. As a result, their performance often degrades under cross-domain scenarios where spoofing attacks, speakers, languages, or recording conditions differ from those observed during training.

To improve the generalization ability of FSD models, a direct and widely adopted solution is multi-dataset training, where spoofing data from diverse sources are aggregated to increase training diversity and coverage~\cite{b7}. Beyond simple data aggregation, existing efforts have improved centralized multi-dataset training from multiple methodological perspectives. At the optimization level, loss-landscape-oriented training strategies~\cite{b8} and their adaptive variants~\cite{b9,b10,b11} have been explored to encourage flatter minima and better cross-domain robustness. From the model-design perspective, pretraining-enhanced architectures and representation-level strategies have been investigated to strengthen feature learning under centralized training~\cite{b12,b13}. Despite their effectiveness, these approaches typically assume access to large and diverse spoofing datasets in a centralized environment, leading to substantial data collection costs, privacy concerns, and limited scalability when spoofing attacks continuously evolve.

As spoofing attacks continue to emerge, FSD models must not only adapt to newly observed fake speech but also preserve previously acquired knowledge. This has motivated continual learning formulations for FSD, often referred to as domain- or attack-incremental learning. Representative approaches include continual domain adaptation~\cite{b14}, adversarial robustness strategies~\cite{b15,b16}, and adaptive sample-selection~\cite{b17} or memory-based strategies~\cite{b18}. However, these continual FSD approaches are primarily developed under a centralized training paradigm. In real-world deployments, speech data are often distributed across institutions, platforms, or edge devices, and different clients may observe highly heterogeneous spoofing sources. Leveraging such distributed data is desirable for broader attack coverage, but raw speech sharing is usually restricted by privacy and ownership constraints. Consequently, how to enable collaboration without sharing raw speech data among distributed clients for continual and generalizable FSD remains an open problem.

To address this challenge, we propose FedCFM, a federated continual domain generalization framework based on Conditional Flow Matching (CFM) for FSD. FedCFM enables clients with heterogeneous spoofing sources to collaboratively learn a generalizable FSD model without sharing raw speech data. Specifically, each client trains a CFM-based generative model to characterize its local spoofing distribution and uploads it to the server, which then redistributes the received generators for embedding-level knowledge sharing. Meanwhile, clients update their FSD classifiers using both local embeddings and synthesized embeddings, and the updated classifier parameters are aggregated on the server to form a global model for the next round. In summary, our main contributions are as follows:

\begin{figure*}[t]
	\centering
	\includegraphics[width=0.8\linewidth]{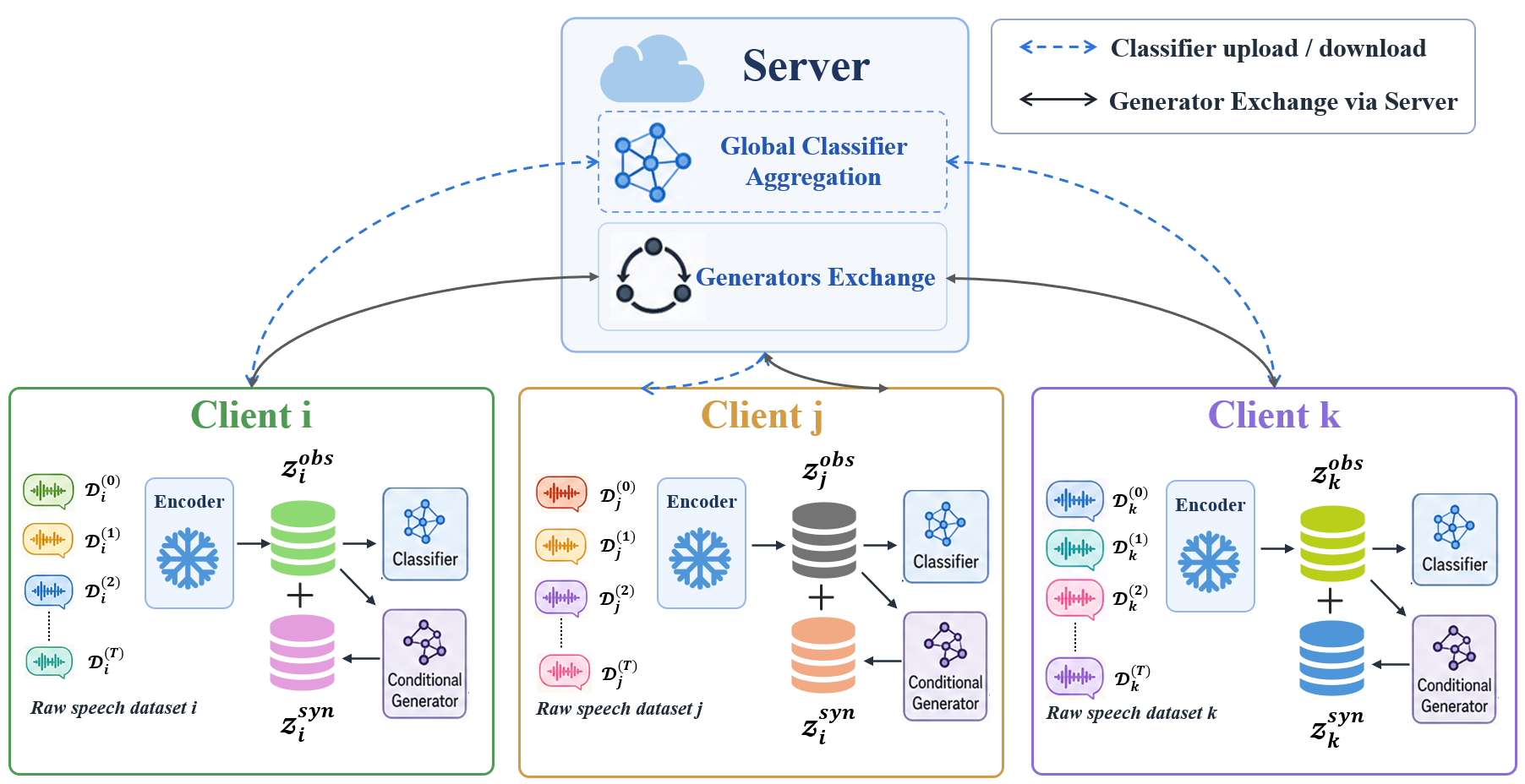}
	\caption{Overview of the proposed FedCFM framework.}
	\label{fig1}
\end{figure*} 

(1) We propose FedCFM, a federated continual domain generalization framework for FSD, where local classifiers are aggregated for global model learning and client-side generative models are exchanged for distribution knowledge sharing.

(2) We equip each client with a CFM-based generator trained using a joint supervised contrastive and flow matching objective to explicitly model spoof-type-specific embedding distributions from locally observed data, jointly enforcing discriminative structure and conditional distribution modeling.

(3) We propose a cross-client generative model exchange mechanism that allows clients to exchange model-level distributional representations rather than raw or directly observable speech data, thereby improving coverage of heterogeneous spoofing conditions and facilitating continual adaptation to emerging attacks.

\section{Related Work}
\subsection{Federated Learning in Deepfake Detection}
Federated learning (FL) has recently gained attention in anti-spoofing and deepfake detection, where privacy-sensitive data are distributed across institutions and devices, making centralization impractical. In fPAD, prior studies have explored federated test-time adaptation~\cite{b19}, split learning under non-IID and unseen domains~\cite{b20}, and domain-invariant feature aggregation~\cite{b21}. In FSD, aggregation-based federated learning has been investigated across data silos~\cite{b22}. These approaches improve cross-domain generalization under privacy constraints but mainly target domain gaps, client heterogeneity, or aggregation design, without addressing continuously emerging spoofing attacks. Generative replay has also been explored in federated continual learning. DGR-FCL~\cite{b32} uses server-side diffusion-based generator to replay, but it targets general class-incremental learning and does not model client-specific spoofing distributions for cross-domain FSD.

Different from these studies, we formulate FSD as a federated continual domain generalization problem. The objective is to improve FSD generalization to unseen and evolving spoofing attacks through cross-client collaboration without sharing raw speech data, by exploiting client-level data heterogeneity and CFM-based spoof-type distribution generation.

\subsection{Federated Domain Generalization}
Federated Domain Generalization (FDG) aims to learn models that generalize to unseen target domains from multiple decentralized and heterogeneous source domains under privacy constraints, where each client has access only to its local data. Existing FDG methods can be broadly categorized by the stage of the Federated Learning (FL) pipeline at which domain heterogeneity is addressed. At the data or feature level, several approaches expand local training distributions through cross-client information sharing, such as FedDG\cite{b23}, CCST\cite{b24}, and GGEUR\cite{b25}. At the server-side aggregation stage, methods like FedFusion\cite{b26} improve global model formation by incorporating distribution-aware aggregation. Representation-level approaches, exemplified by FedCCRL \cite{b27}, enforce cross-client consistency to promote domain-invariant features without centralized data access. More recently, optimization-level strategies have been explored, where FedGaLA \cite{b28} mitigates domain-specific drift through gradient alignment during federated training.

Different from existing FDG methods that mainly rely on feature alignment, representation consistency, or distribution-aware aggregation, our proposed FedCFM explicitly exchanges class-conditional spoofing distributions across clients in the form of client-private generators. By sharing spoof-type-specific generative models rather than features or gradients, FedCFM enables controllable cross-client synthesis of unseen spoofing embeddings and naturally supports continual adaptation.

\begin{algorithm}[t]
	\caption{FedCFM}
	\label{alg1}
	\SetKwInput{KwNotations}{\rmfamily Notations}
	\KwNotations{local classifier training epochs $E_{\text{classifier}}$, local CFM model training epochs $E_{\text{CFM}}$, data arrival interval $M$.
	}
	
	\BlankLine
	Initial global classifier parameters $\phi_G^{0}$
	\For{$t \leftarrow 1$ \KwTo $T$}{
		
		\textbf{CFM updating and redistribution} \hfill (Sec.~A)
		
		\For{$client$ $k=1,\dots,K$ \tcp*[f]{in parallel}}{
			Initial the local generators $G_{\psi_k}^{0}$\\
			\eIf{$t \bmod M = 0$}{
				New speech data arrive at client $k$\;
				Extract embeddings $z_k^{\mathrm{obs}}$ using $f_{\theta_k}$\;
				\For{$e \leftarrow 1$ \KwTo $E_{\text{CFM}}$}{
					Compute $\mathcal{L}_{G_k}(\psi_k^{t-1})$ via Eq.~(3)\;
					$\psi_k^{t} \leftarrow \psi_k^{t-1} - \eta\nabla\mathcal{L}_{G_k}(\psi_k^{t-1})$\;}
				\textbf{Upload} $\psi_k^{t}$ to \textbf{Server}\;
			}{
				$\psi_k^{t} \leftarrow \psi_k^{t-1}$\;
			}
		}
		
		\textbf{Server redistributes} $\{\psi_m^{t}\}_{m \neq k}$ to each client $k$\;
		
		\textbf{Classifier training } \hfill(Sec.~B)
		
		\For{$client$ $k=1,\dots,K$ \tcp*[f]{in parallel}}{
			Initialize $\phi_k^{t} \leftarrow \phi_G^{t-1}$\;
			Replay $z_k^{\mathrm{obs}}$ using $G_{\psi_k}^{t-1}$\;
			Generate $z_k^{\mathrm{syn}}$ using $\{G_{\psi_m}^{t}\}_{m \neq k}$\;
			
			\For{$e \leftarrow 1$ \KwTo $E_\text{classifier}$}{
				Compute $\mathcal{L}_k(\phi_k^{t-1})$ via Eq.~(6)\;
				$\phi_k^{t} \leftarrow \phi_k^{t-1} - \eta \nabla\mathcal{L}_k(\phi_k^{t-1})$\;
			}
			
			\textbf{Upload} $\phi_k^{t}$ to \textbf{Server}\;
		}
		Server obtain $\phi_G^{t}$ according to Eq.~(7). \hfill (Sec.~C)
	}
\end{algorithm}

\begin{table*}[t]
	\centering
	\caption{Generalization performance in terms of EER (\%) of representative domain generalization methods for FSD. Within each block, the best and second-best results per column are highlighted in \textbf{bold} and \underline{underline}, respectively.}
	\label{tab2}
	\setlength{\tabcolsep}{3.6pt}
	
	\begin{tabular}{lccccccccccccc}
		\toprule
		\textbf{Models} & \textbf{Training Set} & \textbf{19LA} & \textbf{CFAD} & \textbf{ASV5} & \textbf{CF} & \textbf{SF} & \textbf{21LA} & \textbf{21DF} & \textbf{ITW} & \textbf{ADD22} & \textbf{AI4T} & \textbf{SONAR} & \textbf{CD-ADD} \\
		
		\midrule
		\multicolumn{14}{l}{\cellcolor{green!12}\textit{Single-dataset Training}} \\
		XLSR-MOE \cite{b35} & \textcircled{1} & 0.74 & – & – & – & – & 2.96 & 2.54 & 12.48 & – & – & – & – \\
		
		VV-SSLD \cite{b36} & \textcircled{1} & 1.91 & 5.88\textsuperscript{$\star$} & 21.44\textsuperscript{$\star$} & 26.51\textsuperscript{$\star$} & 20.44\textsuperscript{$\star$} & 15.92 & 5.67 & 6.10 & 14.46\textsuperscript{$\star$} & 44.19\textsuperscript{$\star$} & 6.51\textsuperscript{$\star$} & 3.55\textsuperscript{$\star$} \\
		
		XLSR-AASIST \cite{b9} & \textcircled{1} & \underline{0.12} & 16.38\textsuperscript{$\star$} & 37.97\textsuperscript{$\star$} & 41.58 & 35.09\textsuperscript{$\star$} & 10.21\textsuperscript{$\star$} & 11.21\textsuperscript{$\star$} & 23.71 & 27.86\textsuperscript{$\star$} & 49.64\textsuperscript{$\star$} & 32.37\textsuperscript{$\star$} & 32.29\textsuperscript{$\star$} \\
		
		Fake-Mamba \cite{b37}  & \textcircled{2} & 10.86\textsuperscript{$\dag$} & \underline{0.74}\textsuperscript{$\dag$} & 23.43\textsuperscript{$\dag$} & 35.66\textsuperscript{$\dag$} & 28.13\textsuperscript{$\dag$} & 17.01\textsuperscript{$\dag$} & 13.64\textsuperscript{$\dag$} & 36.56\textsuperscript{$\dag$} & 22.45\textsuperscript{$\dag$} & 39.93\textsuperscript{$\dag$} & 32.07\textsuperscript{$\dag$} & 40.99\textsuperscript{$\dag$}  \\
		
		XLSR-AASIST \cite{b9} & \textcircled{2} & 2.02\textsuperscript{$\dag$} & \textbf{0.01}\textsuperscript{$\dag$} & 19.76\textsuperscript{$\dag$} & 9.55\textsuperscript{$\dag$} & 10.92\textsuperscript{$\dag$} & 15.28\textsuperscript{$\dag$} & 6.57\textsuperscript{$\dag$} & 6.17\textsuperscript{$\dag$} & 6.19\textsuperscript{$\dag$} & 32.73\textsuperscript{$\dag$} & 8.00\textsuperscript{$\dag$} & 7.67\textsuperscript{$\dag$} \\
		
		WavLM-Nes2Net2 \cite{b38} & \textcircled{3} & 22.53\textsuperscript{$\star$} & 35.84\textsuperscript{$\star$} & \textbf{6.13} & 46.10\textsuperscript{$\star$} & 41.29\textsuperscript{$\star$} & 25.93\textsuperscript{$\star$} & 16.15\textsuperscript{$\star$} & 13.81\textsuperscript{$\star$} & 34.07\textsuperscript{$\star$} & 55.40\textsuperscript{$\star$} & 39.72\textsuperscript{$\star$} & 7.43\textsuperscript{$\star$} \\
		
		XLSR-AASIST \cite{b9} & \textcircled{3} & 17.14\textsuperscript{$\dag$} & 13.01\textsuperscript{$\dag$} & \underline{6.89}\textsuperscript{$\dag$} & 12.58\textsuperscript{$\dag$} & 14.81\textsuperscript{$\dag$} & 14.79\textsuperscript{$\dag$} & 11.38\textsuperscript{$\dag$} & 9.01\textsuperscript{$\dag$} & 12.20\textsuperscript{$\dag$} & 46.76\textsuperscript{$\dag$} & 24.01\textsuperscript{$\dag$} & 1.17\textsuperscript{$\dag$} \\
		
		XLSR-Mamba \cite{b4} & \textcircled{4} & 5.90\textsuperscript{$\dag$} & 24.02\textsuperscript{$\dag$} & 15.82\textsuperscript{$\dag$} & 0.70\textsuperscript{$\dag$} & 8.78\textsuperscript{$\dag$} & 5.31\textsuperscript{$\dag$} & 2.19\textsuperscript{$\dag$} & 11.37\textsuperscript{$\dag$} & 8.66\textsuperscript{$\dag$} & 38.49\textsuperscript{$\dag$} & 7.12\textsuperscript{$\dag$} & 1.45\textsuperscript{$\dag$} \\
		
		XLSR-AASIST \cite{b9} & \textcircled{4} & 3.81 & 11.99\textsuperscript{$\star$} & 17.93\textsuperscript{$\star$} & 0.18 & 7.23\textsuperscript{$\star$} & 8.35\textsuperscript{$\star$} & 2.49\textsuperscript{$\star$} & 9.61 & 5.72\textsuperscript{$\star$} & 28.78\textsuperscript{$\star$} & 8.66\textsuperscript{$\star$} & 1.36\textsuperscript{$\star$} \\

		XLSR-AASIST \cite{b9} & \textcircled{5} & 9.63\textsuperscript{$\dag$} & 8.31\textsuperscript{$\dag$} & 14.82\textsuperscript{$\dag$} & 2.12\textsuperscript{$\dag$} & \textbf{1.00}\textsuperscript{$\dag$} & 18.68\textsuperscript{$\dag$} & 5.82\textsuperscript{$\dag$} & \underline{4.21}\textsuperscript{$\dag$} & 5.43\textsuperscript{$\dag$} & 30.22\textsuperscript{$\dag$} & \underline{1.62}\textsuperscript{$\dag$} & 1.54\textsuperscript{$\dag$} \\
		
		\midrule
		
		\multicolumn{14}{l}{\cellcolor{red!12}\textit{Centralized Multi-dataset Training}} \\
		
		XLSR-AASIST(CSAM) \cite{b9} & \textcircled{1}+\textcircled{4} & 0.31 & 13.33\textsuperscript{$\star$} & 17.51\textsuperscript{$\star$} & \textbf{0.08} & 6.16\textsuperscript{$\star$} & 5.31\textsuperscript{$\star$} & 2.19\textsuperscript{$\star$} & 4.69 & 4.22\textsuperscript{$\star$} & \textbf{23.74}\textsuperscript{$\star$} & 6.10\textsuperscript{$\star$} & 0.71\textsuperscript{$\star$} \\
		
		Wav2DF-TSL \cite{b12} & \textcircled{1}+\textcircled{3}+\textcircled{6} & \textbf{0.10} & – & – & – & – & \textbf{0.87} & \underline{1.95} & 6.83 & – & – & – & – \\
		
		XLSR-ASDG \cite{b39} & \textcircled{1}+\textcircled{7} & – & 2.02 & – & – & – & – & 2.22 & 5.16 & – & – & – & – \\
		WavLM-Nes2Net2\cite{b38} & \textcircled{1}+\textcircled{2}+\textcircled{3}+\textcircled{4}+\textcircled{5} & 0.83\textsuperscript{$\dag$} & 4.47\textsuperscript{$\dag$} & 12.34\textsuperscript{$\dag$} & 1.12\textsuperscript{$\dag$} & 2.76\textsuperscript{$\dag$} & 10.45\textsuperscript{$\dag$} & 4.34\textsuperscript{$\dag$} & 6.41\textsuperscript{$\dag$} & 4.09\textsuperscript{$\dag$} & 27.91\textsuperscript{$\dag$} & 4.16\textsuperscript{$\dag$} & 0.81\textsuperscript{$\dag$} \\
		XLSR-AASIST(CSAM)\cite{b9} & \textcircled{1}+\textcircled{2}+\textcircled{3}+\textcircled{4}+\textcircled{5} & 1.37\textsuperscript{$\dag$} & 3.57\textsuperscript{$\dag$} & 9.88\textsuperscript{$\dag$} & 0.28\textsuperscript{$\dag$} & 2.14\textsuperscript{$\dag$} & 7.74\textsuperscript{$\dag$} & 2.87\textsuperscript{$\dag$} & 5.72\textsuperscript{$\dag$} & \underline{3.77}\textsuperscript{$\dag$} & 27.34\textsuperscript{$\dag$} & 3.82\textsuperscript{$\dag$} & \underline{0.60}\textsuperscript{$\dag$} \\
		
		\midrule
		
		\multicolumn{14}{l}{\cellcolor{blue!12}\textit{Federated Continual Domain Generalization Training}} \\
		
		\textbf{Ours} & \textcircled{1}+\textcircled{2}+\textcircled{3}+\textcircled{4}+\textcircled{5} & 0.48 & 2.98 & 8.16 & \underline{0.10} & \underline{1.91} & \underline{2.23} & \textbf{1.42} & \textbf{3.01} & \textbf{2.85} & \underline{24.46} & \textbf{1.19} & \textbf{0.11} \\
		
		\bottomrule
		
		\\[-0.6em]
		\addlinespace[0.2em]
		
		\multicolumn{14}{l}{\footnotesize
			\shortstack[l]{
				Training datasets: 
				\textcircled{1} 19LA, 
				\textcircled{2} CFAD,
				\textcircled{3} ASVspoof5(ASV5), 
				\textcircled{4} CodecFake(CF),
				\textcircled{5} SpeechFake(SF),
				\textcircled{6} LJSpeech+VCTK, and 
				\textcircled{7} WaveFake+FakeAVCeleb. \\
				\textsuperscript{$\star$} Results are obtained by evaluating officially released pretrained models; \\
				\textsuperscript{$\dag$} Results are obtained by training with the officially released source code on just our dataset setting.\\
				All remaining baseline results are quoted from the corresponding publications.
		}}
		
	\end{tabular}
\end{table*}

\section{Proposed Method}
As shown in Fig. \ref{fig1}, FedCFM follows a standard federated learning protocol consisting of a central server and $K$ distributed clients. Each client $k$ observes speech data from heterogeneous domains over time. The initial local dataset is denoted as $\mathcal{D}_k^{(0)}$, and new data $\Delta \mathcal{D}_k^{(t)}$ become available during communication round $t$. Accordingly, the cumulative dataset up to round $t$ is defined as $\mathcal{D}_k^{(\leq t)} := \mathcal{D}_k^{(0)} \cup (\bigcup_{r = 1}^{t} \Delta \mathcal{D}_k^{(r)})$. Each client $k$ maintains a FSD model composed of a pretrained feature extractor $f_{\theta_k}$ and a classifier $h_{\phi_k}$, together with a client-private CFM-based generative model $G_{\psi_k}$. At each communication round $t$, client $k$ first employs the feature extractor $f_{\theta_k}$ to obtain embeddings $z_k^{\mathrm{obs}}$ from its locally observed speech data $\mathcal{D}_k^{(\leq t)}$ and updates its CFM-based generator $G_{\psi_k}$ to capture spoof-type-specific embedding distributions. Upon local update, the generator is transmitted to the central server, which coordinates the redistribution of client-private generators. Leveraging the received generators, each client synthesizes embeddings $z_k^{\mathrm{syn}}$ corresponding to attack types absent from its local dataset. The classifier $h_{\phi_k}$ is then trained using both locally observed embeddings $z_k^{\mathrm{obs}}$ and synthesized embeddings $z_k^{\mathrm{syn}}$. The updated classifier parameters are uploaded to the server and aggregated via standard federated averaging\cite{b29} to obtain the global model. The above procedure is repeated for $T$ communication rounds.
\subsection{Client-Side Generative Model Based on Conditional Flow Matching}
Given a speech segment, the local feature extractor $f_{\theta}$ maps it to an embedding $z_k^{\mathrm{obs}} \in \mathbb{R}^{L \times D}$, where $L$ and $D$ denote the temporal length and feature dimensionality, respectively.

Each client $k$ trains a lightweight CFM-based generator to model the spoof-type-specific embedding distribution $p_k(z_k^{\mathrm{obs}} \mid c_k)$, where $c_k$ denotes the attack type observed at client $k$. The condition label is encoded as a one-hot vector $e_c \in \mathbb{R}^{C}$, broadcast along the temporal dimension, and concatenated with $z_k^{\mathrm{obs}}$ along the feature dimension to form a condition-aware embedding $\tilde{z}_k^{\mathrm{obs}} \in \mathbb{R}^{L \times (D+C)}$. To improve computational efficiency, we employ a lightweight variant of the original CFM, adapted from Matcha-TTS~\cite{b30}, in which the transformer blocks are replaced by linear-attention modules to reduce model complexity under federated learning constraints.

CFM training follows a flow matching formulation by interpolating between Gaussian noise and the target conditional embedding. Specifically, an interpolation time $s \sim \mathcal{U}(0,1)$ is sampled, and the intermediate state is constructed as $\tilde{z}^{\mathrm{obs}}_{k,s} = \left[1 - (1 - \sigma_{\min}) s \right]\epsilon + s\tilde{z}_k^{\mathrm{obs}}$, where $\epsilon \sim \mathcal{N}(0, I)$ denotes standard Gaussian noise and $\sigma_{\min}$ controls the minimum noise level. The corresponding target velocity is defined as
$u^{\ast} = \tilde{z}_k^{\mathrm{obs}} - (1 - \sigma_{\min}) \epsilon$. The CFM learns a conditional velocity field $v_{\psi_k}$ by minimizing the mean squared error between the predicted and target velocities:
\begin{equation}
	\mathcal{L}_{\mathrm{CFM}}
	=
	\mathbb{E}_{z \sim p_k(z_k^{\mathrm{obs}} \mid c_k),\, \epsilon,\, s}
	\left[
	\lVert v_{\psi_k}(\tilde{z}_{k,s}, s, c_k) - u^{\ast} \rVert_2^{2}
	\right].
\end{equation}
To encourage a more discriminative and condition-structured embedding space across spoofing conditions, a supervised contrastive loss $\mathcal{L}_{\mathrm{supcon}}$~\cite{b31} is applied. This objective promotes intra-condition compactness and inter-condition separability by pulling embeddings associated with the same spoofing condition closer together while pushing those from different conditions farther apart. Such regularization helps the generator capture condition-discriminative distribution patterns and improves the quality of subsequent class-conditional generation. Specifically, $\mathcal{L}_{\mathrm{supcon}}$ is defined as
\begin{equation}
	\mathcal{L}_{\mathrm{supcon}}
	=
	\sum_{i=1}^{N}
	\frac{-1}{|P(i)|}
	\sum_{p \in P(i)}
	\log
	\frac{\exp \left(\mathrm{sim}(\tilde{z}_i, \tilde{z}_p)/\tau\right)}
	{\sum_{a \in A(i)} \exp \left(\mathrm{sim}(\tilde{z}_i, \tilde{z}_a)/\tau\right)},
\end{equation}
where $P(i)$ contains observed and $G_{\psi_k}$-generated embeddings sharing the spoof type of anchor $i$, and $A(i)$ denotes the set of all other samples in the mini-batch. Here, $\mathrm{sim}(\cdot,\cdot)$ represents the cosine similarity between $\ell_2$-normalized embeddings, and $\tau$ is a temperature hyperparameter controlling the concentration of the contrastive distribution. Notably, \(e_c\) conditions the flow network through concatenation with its \(D\)-dimensional features. The resulting \(D+C\) features are projected back to \(D\) for similarity computation, and gradients through the generated representations are propagated to \(\psi_k\). The overall objective for training the CFM-based generator at client $k$ is
\begin{equation}
	\mathcal{L}_{G_{\psi_k}}
	=
	\mathcal{L}_{\mathrm{CFM}}
	+
	\lambda\, \mathcal{L}_{\mathrm{supcon}},
	\label{eq:l_generator}
\end{equation}
where $\lambda > 0$ balances the generative and contrastive terms.

\subsection{Client-Side Continual Training of Classifiers}
After the client-side CFM training described above, each client uploads its locally trained CFM-based generator to the server, which subsequently redistributes the received generators to participating clients in a model-level manner. This cross-client generator exchange enables each client to exploit embedding distribution knowledge learned from other clients. Using the received generators, each client $k$ synthesizes embeddings $z_k^{\mathrm{syn}}$ for attack types that are absent from its local observations, thereby improving the coverage of heterogeneous spoofing conditions.

In addition, due to privacy constraints and limited client storage, historical raw data are not explicitly retained. Instead, client $k$ retains bonafide speech locally and performs generative replay using its CFM-based generator preserved from the previous communication round. The client-private generator is trained to model spoof-type-specific embedding distributions, thereby enabling the synthesis of replay embeddings for previously observed attack types without retaining raw data. A fixed number $N$ of embeddings are generated per attack type to reconstruct historical embeddings $z_k^{\mathrm{obs}}$ for continual training.
Accordingly, the local classifier $h_{\phi_k}$ is trained with $z_k^{\mathrm{obs}} \cup z_k^{\mathrm{syn}}$, using the cross-entropy loss
\begin{equation}
	\mathcal{L}_{\mathrm{task}}
	=
	\mathcal{L}_{\mathrm{CE}} (
	z_k^{\mathrm{obs}} \cup z_k^{\mathrm{syn}}).
\end{equation}

To mitigate catastrophic forgetting, a knowledge distillation loss~\cite{b32} is further introduced to align the updated classifier $h_{\phi_k}^{t}$ with its previous round $h_{\phi_k}^{t-1}$, given by
\begin{equation}
	\mathcal{L}_{\mathrm{KD}} =
	\mathbb{E}_{z^{obs}\in\mathcal{D}_k^{(\leq t)}}
	\left[
	D_{\mathrm{KL}}
	\left(
	h_{\phi_k}^{t-1}(z^{obs})
	\,\Vert\,
	h_{\phi_k}^{t}(z^{obs})
	\right)
	\right].
\end{equation}

Using the two loss terms, the overall classifier objective at client $k$ is defined as
\begin{equation}
	\mathcal{L}_{k}
	=
	\mathcal{L}_{\mathrm{task}}
	+
	\alpha\, \mathcal{L}_{\mathrm{KD}},
\end{equation}
where $\alpha > 0$ balances adaptation to newly emerging spoofing attacks and preservation of previously learned knowledge.

\subsection{Server-Side Coordination and Model Aggregation}
The server in FedCFM is responsible for global aggregation of local classifiers and coordinated exchange of generative models across clients. Specifically, at each communication round $t$, the server aggregates local classifier parameters using equal-weight averaging:
\begin{equation}
	h_{\phi_\mathrm{global}}^{t}
	=
	\frac{1}{K}
	\sum_{k=1}^{K}
	h_{\phi_k}^{t},
	\label{eq:fedavg}
\end{equation}
where $K$ denotes the number of participating clients. 

Then, the aggregated parameters are broadcast for the next round of local training. Every \(M\) rounds, the server redistributes the updated generators to the other clients. This mechanism broadens the coverage of heterogeneous spoofing conditions without requiring direct sharing of raw speech signals. The overall procedure is summarized in Algorithm\ref{alg1}.

\subsection{Data locality and Privacy claim}
Our proposed FedCFM framework enables collaborative learning while keeping raw speech and XLSR embeddings local. Each client extracts 5th-layer XLSR embeddings and trains a CFM-based generator to model spoof-type-specific distributions; prior studies show that waveform reconstruction from such intermediate representations typically requires dedicated inversion or synthesis modules~\cite{b33,b34}. During federated training, the server aggregates classifier parameters and redistributes generators, enabling distribution-level knowledge transfer without transmitting raw speech or extracted representations, thereby mitigating privacy risks.

\subsection{Communication Cost}
FedCFM exchanges classifier parameters every round and CFM generators every \(M=10\) rounds. The classifier and generator contain \(0.45\)M and \(10.03\)M parameters, respectively. Each client uploads its classifier and downloads the aggregated classifier per round; in generator-exchange rounds, it additionally uploads its local generator and downloads the other \(K-1\) generators. With \(K=4\) and FP32, the per-client communication overhead is approximately \(3.58\)~MB in regular rounds and \(164.1\)~MB in generator-exchange rounds, including classifier communication. Over \(T=70\) rounds, this yields an average overhead of \(19.6\)~MB per client per round.

\begin{table*}[t] 
	\centering 
	\caption{Comparison with advanced FDG methods in terms of EER (\%)} 
	\label{tab3} 
	\setlength{\tabcolsep}{3.8pt}
	\begin{tabular}{p{2cm}@{\hspace{1.8pt}}cccccccccccc} 
		\toprule \textbf{Method} & \textbf{19LA} & \textbf{CFAD} & \textbf{ASVspoof5} & \textbf{Codecfake} & \textbf{SpeechFake} & \textbf{21LA} & \textbf{21DF} & \textbf{ITW} & \textbf{ADD22} & \textbf{AI4T} & \textbf{SONAR} & \textbf{CD-ADD}\\ 
		\midrule GGEUR \cite{b25} & 0.67 & 3.20 & 11.92 & 3.77 & 5.13 & 3.59 & 2.72 & 5.67 & 3.59 & 26.65 & 1.90 & 0.33\\ 
		FedFusion \cite{b26} & 2.55 & 7.80 & 18.34 & 6.47 & 10.22 & 8.58 & 6.02 & 6.63 & 4.37 & 44.60 & 2.44 & 3.16 \\ 
		FedCCRL \cite{b27} & 0.98 & 4.32 & 13.20 & 5.29 & 7.35 & 5.14 & 3.76 & 5.11 & 3.99 & 36.34 & 2.00 & 0.98 \\ 
		FedGaLA \cite{b28} & 1.32 & 5.16 & 15.69 & 5.90 & 8.44 & 7.56 & 4.94 & 5.82 & 4.07 & 39.43 & 2.11 & 1.93\\ 
		\midrule 
		\textbf{Ours} & \textbf{0.48} & \textbf{2.98} & \textbf{8.16} & \textbf{0.10} & \textbf{1.91} & \textbf{2.23} & \textbf{1.42} & \textbf{3.01} & \textbf{2.85} & \textbf{24.46} & \textbf{1.19} & \textbf{0.11} \\
		\bottomrule 
	\end{tabular} 
\end{table*}

\begin{table*}[t] 
	\centering 
	\caption{Ablation study of the FedCFM framework in terms of EER (\%).} 
	\label{tab4} 
	\setlength{\tabcolsep}{4pt} 
	\begin{tabular}{lcccccccccccc}
		\toprule
		\textbf{Method} & \textbf{19LA} & \textbf{CFAD} & \textbf{ASVspoof5} & \textbf{Codecfake} & \textbf{Speechfake} & \textbf{21LA} & \textbf{21DF} & \textbf{ITW} & \textbf{ADD22} & \textbf{AI4T} & \textbf{SONAR} & \textbf{CD-ADD} \\
		\midrule
		w/o CFM Upload & 3.97 & 10.08 & 15.26 & 0.77 & 3.83 & 5.72 & 4.91 & 6.50 & 6.34 & 32.95 & 4.68 & 3.60 \\
		w/o $\mathcal{L}_{supcon}$ & 1.85 & 4.77 & 10.51 & 0.51 & 3.13 & 3.60 & 2.79 & 4.38 & 4.22 & 29.83 & 2.56 & 1.48 \\
		w/o $\mathcal{L}_{KD}$ & 1.72 & 4.54 & 9.16 & 0.44 & 2.56 & 3.15 & 2.24 & 4.19 & 3.66 & 28.15 & 2.22 & 1.34 \\
		\textbf{Ours} & \textbf{0.48} & \textbf{2.98} & \textbf{8.16} & \textbf{0.10} & \textbf{1.91} & \textbf{2.23} & \textbf{1.42} & \textbf{3.01} & \textbf{2.85} & \textbf{24.46} & \textbf{1.19} & \textbf{0.11} \\
		\bottomrule
	\end{tabular}
\end{table*}

\begin{table*}[t]
	\centering
	\caption{Comparison of Generative Modeling Strategies in FedCFM in terms of EER (\%)}
	\label{tab5}
	\setlength{\tabcolsep}{4pt}
	\begin{tabular}{p{1.7cm}@{\hspace{1.8pt}}cccccccccccccc}
		\toprule
		\textbf{Generator} 	& \textbf{\#Params}	& \textbf{19LA} & \textbf{CFAD} & \textbf{ASV5} & \textbf{Codecfake} & \textbf{SpeechFake} & \textbf{21LA} & \textbf{21DF} & \textbf{ITW} & \textbf{ADD22} & \textbf{AI4T} & \textbf{SONAR} & \textbf{CD-ADD} \\
		\midrule
		VAE        & 10.89M & 5.18 & 8.10 & 13.94 & 2.57 & 4.77  & 4.53 & 3.07 & 5.89 & 5.41 & 30.72 & 2.63 & 0.76 \\
		Diffusion  & 8.18M & 1.29 & 3.32 & 9.03 & 0.19 & 3.55 & 3.49 & 1.98 & 4.05 & 3.27 & 29.18 & 1.52 & 0.44 \\
		\textbf{Ours}(Flow)      & 10.03M & \textbf{0.48} & \textbf{2.98} & \textbf{8.16} & \textbf{0.10} & \textbf{1.91} & \textbf{2.23} & \textbf{1.42} & \textbf{3.01} & \textbf{2.85} & \textbf{24.46} & \textbf{1.19} & \textbf{0.11}\\
		\bottomrule
	\end{tabular}
\end{table*}

\section{Experiments and Analysis}
\subsection{Datasets}

We train and evaluate FedCFM in a federated continual FDG setting with four clients, each initialized with a distinct FSD dataset to emulate realistic domain shifts across clients.

\textbf{ASVspoof2019LA}~\cite{b40}: a standard English benchmark for TTS/VC spoofing detection, with six known attacks in training/development and unseen attacks in evaluation for cross-attack generalization. We use the original test set.

\textbf{Codecfake}~\cite{b9}: a large-scale bilingual dataset for ALM-based deepfake detection, covering seven neural audio codecs over English and Chinese speech. We evaluate on test conditions C1--C7.

\textbf{CFAD}~\cite{b41}: a Chinese fake audio detection dataset with twelve fake-audio types, including eleven synthesis techniques and one partially fake type. We use the clean evaluation split covering all fake types.

\textbf{ASVspoof5}~\cite{b42}: a large-scale crowdsourced English benchmark for spoofing, deepfake, and adversarial attack detection, covering diverse acoustic conditions and stronger TTS/VC and adversarial attacks. We use the original test set.

To simulate emerging attacks, we select 28 spoofing types from SpeechFake~\cite{b43} and partition them into four disjoint subsets of seven types, one per client. Each client receives one previously unseen type at the start of each of the seven incremental stages, with each stage lasting $M=10$ communication rounds.

We further evaluate the FSD model obtained by FedCFM on seven out-of-domain test sets, which are stated as follows.

\textbf{ASVspoof2021LA}~\cite{b44}: an out-of-domain LA benchmark with TTS/VC attacks and codec/transmission effects.

\textbf{ASVspoof2021DF}~\cite{b44}: a benchmark for compressed, source-mismatched online speech, targeting real-world deepfake dissemination.

\textbf{In-the-Wild (ITW)}~\cite{b45}: a real-world deepfake dataset from online sources, covering bona fide and spoofed public-figure speech.

\textbf{ADD2022}~\cite{b46}: a deep synthesis detection dataset covering low-quality, partially fake, and game-scenario audio.

\textbf{AI4T}~\cite{b47}: a multilingual real-world deepfake dataset from social media, covering misinformation, scam, and entertainment.

\textbf{SONAR}~\cite{b48}: a synthetic AI-audio detection benchmark from diverse synthesis platforms and TTS systems.

\textbf{CD-ADD}~\cite{b49}: a cross-domain deepfake dataset from zero-shot TTS, with noise, compression, and codec perturbations.

\subsection{Implementation Details}

We instantiate the framework with \(K=4\) clients, each integrating an XLSR-300M feature extractor~\cite{b50}, an AASIST classifier~\cite{b1}, and a CFM-based generator adapted from Matcha-TTS~\cite{b30} with linear attention to improve parameter efficiency. All waveforms are truncated to their first 64,600 samples or repeated and then truncated to this length. Fifth-layer XLSR embeddings are used to train the classifier and generator. Classifiers share a common initialization, while generators are initially trained locally from scratch.

Training proceeds for \(T=70\) communication rounds, partitioned into seven incremental stages, each comprising \(M=10\) rounds, with \(10\) local epochs per client per round. Classifier parameters are uploaded and aggregated every communication round. Generators are retrained and exchanged every \(M\) rounds. For generative replay, we synthesize \(N=2{,}500\) embeddings per spoofing type. The CFM generator uses a minimum noise level of \(\sigma_{\min}=10^{-6}\) and \(10\) numerical integration steps.

All trainable models are optimized using Adam with a learning rate of \(10^{-5}\). Batch sizes are \(512\) for the classifier and \(64\) for the CFM generator. The distillation and contrastive loss weights are \(\alpha=0.3\) and \(\lambda=0.7\), respectively, with a contrastive temperature of \(\tau=0.07\).

\subsection{Results and Analysis}
\subsubsection{Comparison with Representative FSD Methods}
Table~\ref{tab2} compares the EER results of our FedCFM-based FSD model with representative FSD methods. From Table~\ref{tab2}, although single-dataset training methods perform well on their in-domain datasets, they degrade markedly on out-of-domain datasets. In contrast, our FedCFM-based FSD model maintains competitive in-domain performance while consistently achieving lower EER on out-of-domain datasets, demonstrating stronger cross-domain generalization.

Compared with centralized multi-dataset training based on static aggregation, FedCFM consistently achieves superior performance under identical dataset combinations, which can be attributed to its distribution-level modeling mechanism rather than naive data pooling. While centralized training optimizes a fixed empirical distribution determined by the proportions of raw data from different datasets, such a strategy remains strongly influenced by dataset imbalance and dominant domains. In contrast, FedCFM explicitly reshapes the training distribution through cross-client generator exchange and class-balanced generative replay. By learning and sampling from class-conditional embedding distributions across clients, the framework mitigates domain dominance, improves coverage of underrepresented spoof types, and expands the effective support of the training distribution, thereby promoting stronger cross-domain generalization.

\subsubsection{Comparison with Advanced FDG Methods}
As shown in Table~\ref{tab3}, the proposed FedCFM-based FSD model consistently achieves the lowest EER among the four FDG-based baselines. This improvement may primarily arise from the explicit modeling and exchange of spoof-type-specific generative models in the embedding space, enabling clients to leverage shared global spoofing knowledge rather than relying solely on implicit alignment or parameter aggregation.

\subsubsection{Ablation Study}

Table~\ref{tab4} reports the ablation results of key components in the FedCFM framework. As shown in Table~\ref{tab4}, the "w/o CFM Upload" variant removes cross-client generator exchange while retaining local balanced generative replay, causing a substantial performance drop across all datasets, which demonstrates the importance of cross-client generative model exchange mechanism. We can also see that removing $\mathcal{L}_{\mathrm{supcon}}$ degrades performance, indicating its role in enhancing feature discrimination during generative modeling. Removing $\mathcal{L}_{\mathrm{KD}}$ also leads to performance decline, highlighting its importance in preserving previously learned knowledge under continual and heterogeneous training conditions.

\subsubsection{Effect of Different Generative Models}

Table~\ref{tab5} compares the EER results of FSD models using different generative backbones in FedCFM, including VAE-based~\cite{b51}, diffusion-based~\cite{b52}, and our CFM-based models. As illustrated in Table~\ref{tab5}, our proposed CFM-based generator consistently yields the lowest EER across all test datasets, which may be attributed to the ability of CFM to directly model conditional feature transformations through deterministic and continuous trajectories, leading to more accurate structure-preserving feature synthesis.

\section{Conclusion}
We propose FedCFM, a federated continual domain generalization framework based on Conditional Flow Matching (CFM) for Fake Speech Detection (FSD). By exchanging client-private generators and integrating generative replay with knowledge distillation, FedCFM adapts to emerging attacks while mitigating forgetting. Under the same training configuration, experiments on five in-domain and seven out-of-domain test sets show that FedCFM consistently improves Equal Error Rate (EER) over the evaluated centralized and federated baselines, demonstrating strong generalization.

\end{document}